\documentclass[pra,twocolumn,superscriptaddress,nobibnotes]{revtex4-2}
\usepackage[utf8]{inputenc}
\usepackage{amsmath,bm}
\usepackage{amsfonts}
\usepackage{amssymb}
\usepackage{xcolor}
\usepackage{braket}
\usepackage{graphicx}
\usepackage{subfigure}
\usepackage{todonotes}
\usepackage{subfigure}
\usepackage[colorlinks=true, allcolors=blue]{hyperref}
\usepackage{dsfont}

\begin{document}
	
	\title{Surface plasmon polaritons with extended  lifetime}
	
	\author{Rasim Volga Ovali}
	\affiliation{Recep Tayyip Erdogan University, 53100 Rize, Turkey}
	\author{Mehmet Emre Tasgin}
	\affiliation{Institute of Nuclear Sciences, Hacettepe University, 06800 Ankara, Turkey}
	
	\date{\today}
	
	\begin{abstract}
The propagation distance of surface plasmon polaritons (SPPs) on metal nanowires is severely limited by their short lifetime, primarily due to strong metallic losses. In this work, we show that the lifetime—and thus the propagation distance—of SPPs can be significantly extended through the use of Fano resonances. Our FDTD simulations demonstrate that the SPP intensity at a fixed propagation distance can be enhanced by approximately 30 times. Furthermore, this enhancement factor is multiplicative with improvements achieved through other methods. We emphasize that this result represents only a starting point, as no optimization was performed due to limited computational resources.
	\end{abstract}
	
	\maketitle

Incident electromagnetic radiation can excite near-field oscillations on the surfaces of metal nanowires~(MNWs). These excitations, surface plasmon polaritons~(SPPs)~\cite{raether2006surface}, propagate along the MNW surface as collective charge oscillations. They can be excited, for instance, by a dipole positioned near the NW edge~\cite{akimov2007generation}, see Fig.~1, or using Kretschmann and similar geometries~\cite{vinogradov2018exciting}. SPPs on NWs behave as the nanoscale analogs of  fiberoptic cables as they operate at optical frequencies ---providing a large bandwidth--- and possesses propagation speeds close to the speed of light. These make utilization of SPPs and MNWs as nanoscale fiberoptic cables in electronic chips is a long-standing objective~\cite{ekmelScience}. The current interconnects ---carry the processed data among different components of the CPU--- employ electronic data transfer. SPPs have the potential to provide thousands of times larger data transfer rates on the same NW architecture~\cite{kobrinsky2004chip,barnes2003surface,maier2005plasmonics}. This is like comparing a fiber-internet with an ADSL. In current CPUs the processing speed is limited by the poor electronic transfer rates of the interconnects~\cite{kobrinsky2004chip}.

Despite the tens of km propagation lengths of fiberoptic pulses, however, SPP propagation lengths are below $\mu$m due to the strong metallic absorption~\cite{verhagen2009nanowire} \footnote{Long-range SPPs in dielectric-metal-dielectric structures propagation distance can reach even cm. However, confinement in these structures is over microns that limits the their use regarding the miniaturization.}. This is one of the two major problems avoiding an effective SPP data transfer over  interconnects. The second problem is the immaturity of the nanoscale integrated lasers, spaser~(surface plasmon laser)~\cite{kewes2017limitations,azzam2020ten}, that can trigger coherent SPP excitations. Yet, one should not directly compare the fiberoptic propagation length with the  SPP propagation distance regarding their technological use. This is because, fiberoptics is employed for intercontinental data transfer while  tens of micrometer propagation lengths could be sufficient for SPPs. The latter is to be utilized in miniature microchips. Therefore, inventing a solution for extending the SPP propagation distance ---that multiplies the enhancement obtained by other techniques---  possesses great importance.

	\begin{figure}
	\centering
	\includegraphics[width=1\linewidth]{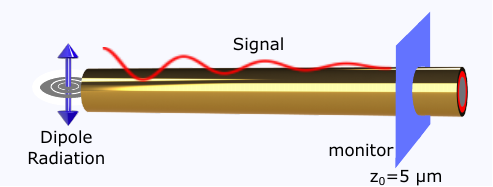}
	\caption{We investigate SPP propagation over a NW. The SPP oscillations are excited by a dipole polarized perpendicular to the NW axis. We investigate the mean $E$-field intensity on a monitor~(blue) positioned at $z=z_0=5\: \mu{\rm m}$. We compare the two NW configurations presented in Figs.~2a and 2b. 
	}
	\label{fig1}
	\end{figure} 

Here, we come up with a method that is not only compatible with other propagation enhancement techniques, but also (indeed) multiplies them. In this study, we clearly show that a Fano resonance, induced by a quantum emitter~(QE) layer~\cite{leng2018strong,wu2010quantum,shah2013ultrafast,tacsgin2018fano,limonov2017fano}, can multiply the SPP propagation intensity by an enhancement factor~(EF) of ${\rm EF_{fano}}\sim 30$. The QE-layer, comprised of densely-implanted defect-centers~\cite{murata2011high,rotem2007enhanced,ngandeu2024hot}, increases the lifetime of SPPs. We demonstrate the effect via exact solutions of the 3D Maxwell equations using FDTD package of a reliable software Lumerical~\cite{lumerical}.

	\begin{figure}
	\centering
	\includegraphics[width=1\linewidth]{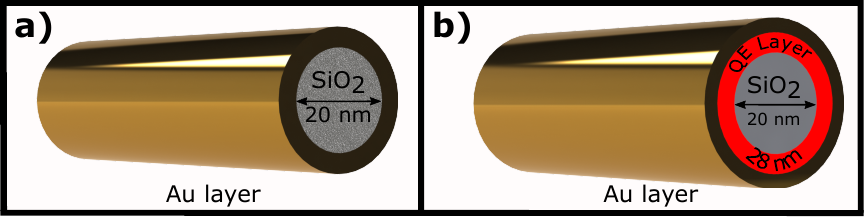}
	\caption{We compare  SPP propagation in two different NW configurations: a NW structure in the (a) presence, Fig.~2a, and (b) absence, Fig.~2b, of a quantum emitter~(QE)-layer. The QE-layer is made of a densely-implanted defect-centers~\cite{murata2011high,rotem2007enhanced,ngandeu2024hot} and introduces a Fano resonance~(dip) in Fig.~3b. We use a ${\rm SiO_2}$-core as such structures are known to possess smaller plasmon decay rates~\cite{rajput2020forster,issah2023epsilon}. 
	}
	\label{fig1}
	\end{figure}

We couple a dipole to a NW and investigate the SPP propagation along the NW. See Fig.~1. We compare the propagation in the (i) absence, Fig.~2a, and (ii) presence of the QE-layer, Fig.~2b. We record the time evolution of the electric field $E({\bf r},t)$ on a plane monitor~(blue in Fig.~1) placed at a $z=z_0=$5$\:\mu{\rm m}$ distance for configurations (i) and (ii). We use a ${\rm SiO_2}$~(insulator) core in both cases. We compare the the mean electric field intensity on the monitor and observe the enhancement factor ${\rm EF_{fano}}\sim30$. Compare Figs.~4a and 4b. We note that SPP propagation in a ${\rm SiO_2}$-core NW~(Fig.~2a) is already enhanced ${\rm EF}_{\scriptscriptstyle \rm SiO_2}=$15 times compared to a bare gold NW (not shown) via employing another technique~\cite{rajput2020forster,issah2023epsilon,adato2010radiative}. In these studies~\cite{rajput2020forster,issah2023epsilon} presence of the ${\rm SiO_2}$-core enhances effective lifetime of the localized SP~(LSP) modes, because losses are compensated via optical gain incorporated in the insulator-core. The Fano enhancement factor~${\rm EF_{fano}}$ multiplies the enhancement factor obtained via the other technique, i.e., ${\rm EF_{tot}}= {\rm EF_{fano}} \times {\rm EF_{\scriptscriptstyle \rm SiO_2}}$. This is our primary finding in this study.

{\it Fano resonances} appear when a bright plasmon mode is coupled to a  narrow band quantum emitter~(QE)~\cite{leng2018strong,wu2010quantum,shah2013ultrafast,tacsgin2018fano,limonov2017fano}~\cite{PSDarkmodes}. FR is the plasmon analog of electromagnetically-induced transparency~(EIT)~\cite{fleischhauer2005electromagnetically} studied in atomic vapors. Presence of the QE introduces two absorption/emission paths that operate destructively~\cite{alzar2001classical}. An excitation dip~(compare Figs. ~3a and 3b) appears at the QE level-spacing $\omega_{\rm f}\simeq \Omega_{\rm \scriptscriptstyle QE}$. The plasmonic excitation ---normally at its resonance for $\omega=\omega_{\rm f}$--- becomes transparent or displays a dip at $\omega_{\rm f}$.

In the present study, SPP spectrum of the gold-coated ${\rm SiO_2}$-core NW~(Fig.~2a) exhibits a broad resonance at $\Omega_{p}=$545 nm, see Fig.~3a. In current CPUs the processing speed is limited by the poor electronic data transfer rates of the interconnects~\cite{kobrinsky2004chip}. In Fig.~2b, we use a QE-layer coating in between the gold-layer and the ${\rm SiO_2}$-core. We choose the QE's level-spacing at the SPP resonance, i.e., $\Omega_{\rm \scriptscriptstyle QE}=\Omega_p=$545 nm, in order to achieve a strong plasmon-QE interaction. QE-layer induces a Fano dip~(Fig.~3b) in the SPP spectrum at $\omega_{\rm f}=$550 nm which is close to $\Omega_{\rm \scriptscriptstyle QE}=$545 nm. FR, $\omega_{\rm f}$, does not appear exactly on $\Omega_{\rm \scriptscriptstyle QE}$ because of the retardation effects~\cite{PSwf}.

A FR exhibits two different~(actually opposite) effects in the ({\it 1}) time evolution  and in the ({\it 2}) steady-state. When the coupled plasmon-QE system is excited by a  pulsed laser, ({\it 1}), one observes that lifetime of plasmon oscillations extends compared to a bare plasmonic system~\cite{PSgoldNW}. ({\it 2}) A contrary behavior is observed when the coupled system is driven continuously via a CW laser. That is, when the steady-state is reached, a Fano transparency~(suppression) appears, e.g., in Fig.~3b~\cite{PSsteadystate}. 

	\begin{figure}
	\centering
	\includegraphics[width=1\linewidth]{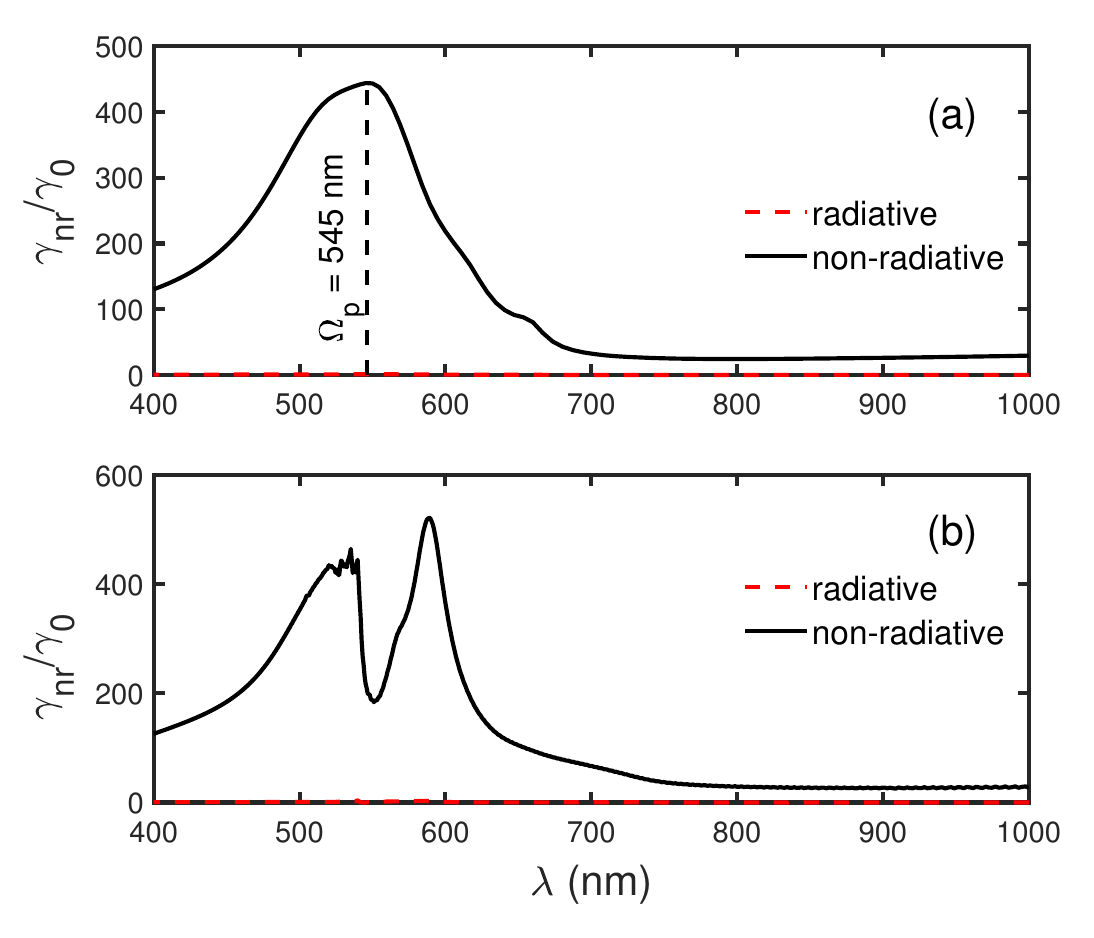}
	\caption{Radiative~($\gamma_{\rm r}$) and nonradiative~($\gamma_{\rm nr}$) decay rates when the QE-layer is (a) absent and (b) present. See Figs.~2a and 2b, respectively. In (b), QE-layer introduces a Fano resonance~(a dip) at $\omega_{\rm f}=550$ nm. In the SPP propagation simulations, Fig.~4, we choose the carrier frequency of the dipole at $\omega_{\rm dip}=\omega_{\rm f}=$550 nm to achieve larger lifetime enhancements.
	}
	\label{fig3}
	\end{figure}

Actually, such a lifetime extension phenomenon has already been observed for localized~(non-propagating) surface plasmons~\cite{yildiz2020plasmon,ovali2021single,sun2023correlation,qin2024tailoring}  and is responsible for the linear and nonlinear Fano enhancements with pulsed lasers~\cite{limonov2017fano}. Lifetime extension for SPPs, however, has not been demonstrated before, to our best knowledge. Such an enhancement has the potential to open the door for major improvements on chip-based technologies.

{\it FDTD Simulations}.--- We compare the mean electric-field intensity on the plane monitor~(blue in Fig.~1) for the NW configurations depicted in (i) Fig.~2a and (ii) Fig.~2b.
We employ the FDTD package of the reliable toolbox Lumerical~\cite{lumerical} and use the actual~(experimental) dielectric functions for gold and ${\rm SiO_2}$. We position a dipole near the NW and first calculate the radiative~($\gamma_{\rm r}$) and nonradiative~($\gamma_{\rm nr}$) decay rates as plotted in Figs.~3a and 3b. In both cases we observe that dipole radiation is mainly transferred into the NW: $\gamma_{\rm r}$ are very small compared to $\gamma_{\rm nr}$. In Fig.~3a, we observe that the SPP resonance of the ${\rm SiO_2}$-gold system~(Fig.~2a) is centered at $\Omega_p=$545 nm.

In order to obtain a strong  QE-NW coupling, we choose the level-spacing of the QE-layer at $\Omega_{\rm \scriptscriptstyle QE}=\Omega_p=$545 nm. In Fig. 3b, we observe that presence of the QE-layer introduces a Fano dip at $\omega_{\rm f}=550$ nm. Retardation effects shift the Fano resonance to  $\omega_{\rm f}=550$ nm~\cite{PSwf}.  
We simulate the QE-layer~(defect-centers) by a Lorentzian dielectric function~\cite{wu2010quantum,shah2013ultrafast,leng2018strong} of resonance $\Omega_{\rm \scriptscriptstyle QE}=$545 nm, linewidth $\gamma_{\rm \scriptscriptstyle QE}=10^{10}$ Hz and oscillator strength $f_{\rm osc}=0.2$~\cite{wu2010quantum,shah2013ultrafast,leng2018strong,murata2011high,rotem2007enhanced,ngandeu2024hot}. 

 Next, we investigate the time evolution of the SPP propagation for the two NW configurations. SPP excitation is induced by the dipole source. We choose the carrier frequency of the dipole source at $\omega_{\rm dip}=\omega_{\rm f}$ in order to achieve larger SPP lifetimes.  
 We record the electric field square $|E({\bf r},t)|^2$ data at each ${\bf r}=[x,\: y,\: z_0]$ point on the monitor~(blue in Fig.~1), with $z_0=$5 $\mu$m.
We  calculate the mean-$E$-field intensity at each points by integrating in the time domain, i.e., $I(x,y)=\int dt \: |E(x,y,z_0,t)|^2$.

 In Figs.~4a and 4b, we plot mean electric-field intensity $I(x,y)$ for the two NW configurations presented in Fig.~2a and 2b, respectively. We clearly observe that presence of the QE-layer increases the mean-$E$-intensity at $z_0=5\: \mu{\rm m}$ about 30 times compared to Fig.~2a. 
We note that this enhancement factor is  obtained even without optimizing the parameters related to interaction strength and material dimensions, so the SPP modes. That is, much larger enhancements should be available via optimization~\cite{PSoptimization}. 

\begin{figure}
	\centering
	\includegraphics[width=\linewidth,trim=0cm 0cm 3cm 0cm,clip]{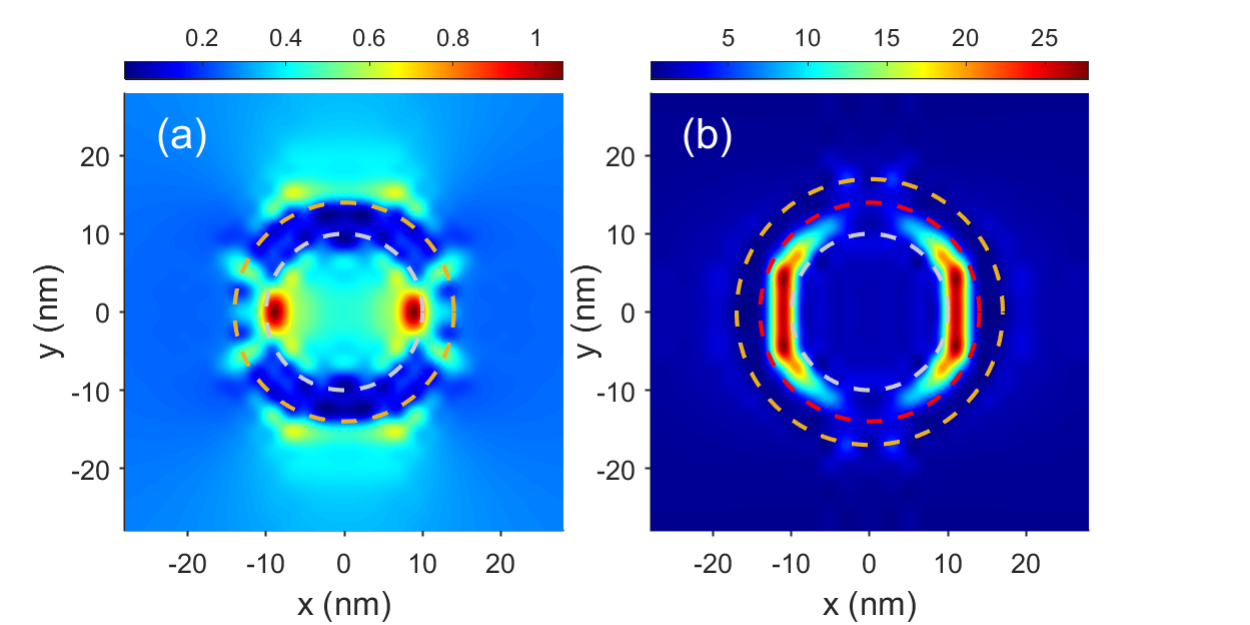}
	\caption{Mean electric-field intensity distribution on the plane monitor~(blue in Fig.~1) in the (a) absence, Fig.~2a, and (b) presence of the QE-layer. The intensity at $z=z_0=5 \: \mu{\rm m}$ is enhanced about 30 times. This enhancement factor ${\rm EF_{fano}}\sim 30$ multiplies the enhancement factor coming from the other enhancement technique~(use of ${\rm SiO_2}$ insulator core~\cite{rajput2020forster,issah2023epsilon}) ${\rm EF_{SiO_2}}=15$. That is, the total EF is ${\rm EF_{tot}} ={\rm EF_{fano} } \times {\rm EF_{SiO_2}}=$ 450. 
	}
	\label{fig4}
\end{figure} 

We use a ${\rm SiO_2}$-core NW structure because such a structure has been demonstrated to possess smaller plasmon decay rates in the LSP case~\cite{rajput2020forster,issah2023epsilon}.
We also performed FDTD simulations with a bare gold NW~(not presented). ${\rm SiO_2}$-core NW in Fig.~2a results an ${\rm EF_{\rm \scriptscriptstyle SiO_2}}=$15 times intensity enhancement factor on the plane monitor compared to a bare gold NW. The enhancement factor ${\rm EF_{fano}}=$30, appears solely due to Fano resonance, multiplies the  enhancement factor ${\rm EF_{\rm \scriptscriptstyle SiO_2}}$. That is, final enhancement is ${\rm EF_{tot}}= {\rm EF_{fano}}  \times {\rm EF_{\rm \scriptscriptstyle SiO_2}}=$450. The two enhancement methods are compatible. We should note that the ${\rm SiO_2}$-gold NW we use in our simulations~(Fig.~2b) is not optimized for parameters like gold vs ${\rm SiO_2}$ thicknesses, different carrier frequencies $\omega_{\rm dip}$ and linewidths.
 Refs.~\cite{rajput2020forster,issah2023epsilon} show that substantially smaller plasmon decay rates appear for particular choices of these parameters~\cite{PSgoldNW} where epsilon-near-zero~(ENZ) like effects appear~\cite{issah2023epsilon}.

In \textit{conclusion}, exact solutions of 3D Maxwell equations demonstrate that lifetime of SPPs can be extended substantially when Fano resonances are employed. The demonstrated phenomenon is of crucial importance in microchip technologies. This is because, data transfer rates with SPPs on current NW interconnect architecture is thousands of times faster than the electronic rates. Our method makes this objective ---utilization of SPPs on NWs as nanoscale fiberoptic cables--- one step closer as it comes as a multiplicative factor. That is, our method is not an alternative to other techniques but multiplies them.

Finally, we would like to raise the following important point. We unfortunately could perform the simulations only for a few sets of system parameters such as layer thickness, QE oscillator strength and level-spacing, and pulse center and width~\cite{PSoptimization}. For instance, in Ref.~\cite{yildiz2020plasmon}, it is underlined that maximum LSP lifetime enhancement~(almost 10 times) takes place at a critical coupling. In Ref.~\cite{yildiz2020plasmon}, a $10^{-13}$ sec lifetime dark mode is utilized. Here, we use a longer lifetime ($10^{-10}$ sec) QE-layer.  
 Thus, one should expect much larger  ${\rm EF_{fano}}$ and $ {\rm EF_{\rm \scriptscriptstyle SiO_2}}$~\cite{PSgoldNW} enhancement factors via optimization process.

				\bibliography{bibliography}	
			\end{document}